# Frequency landscape of tip-growing plants

Mariusz Pietruszka[*] and Aleksandra Haduch-Sendecka

Laboratory of Plant Physiology, Faculty of Biology and Environment Protection, University of Silesia, Katowice, Poland

*) Corresponding author: mariusz.pietruszka@us.edu.pl (M. Pietruszka)

**Abstract**

It has been interesting that nearly all of the ion activities that have been analysed thus far have exhibited oscillations that are tightly coupled to growth. Here, we present discrete Fourier transform (DFT) spectra with a finite sampling of tip-growing cells and organs that were obtained from voltage measurements of the elongating coleoptiles of maize *in situ*. The electromotive force (EMF) oscillations (~ 0.1 µV) were measured in a simple but highly sensitive RL circuit, in which the solenoid was initially placed at the tip of the specimen and then was moved thus changing its position in relation to growth (EMF can be measured first at the tip, then at the sub-apical part and finally at the shank). The influx- and efflux-induced oscillations of $Ca^{2+}$, along with $H^+$, $K^+$ and $Cl^-$ were densely sampled (preserving the Nyquist theorem in order to 'grasp the structure' of the pulse), the logarithmic amplitude of pulse spectrum was calculated, and the detected frequencies, which displayed a periodic sequence of pulses, were compared with the literature data. A band of life vital individual pulses was obtained in a single run of the experiment, which not only allowed the fundamental frequencies (and intensities of the processes) to be determined but also permitted the phase relations of the various transport processes in the plasma membrane and tonoplast to be established. A discrete frequency spectrum (like the hydrogen spectrum in quantum physics) was achieved for a growing plant for the first time, while all of the metabolic and enzymatic functions of the life cell cycle were preserved using this totally non-invasive treatment.

**Keywords** calcium; chloride; electromotive force; Fourier analysis; growth; hydrogen; ion transport; oscillations; potassium

**Introduction**

The physico-chemical nature of wall loosening in growing plants is still an issue that is being investigated (Breidwood et al., 2014). Oscillations appear to be generic processes in biological systems in the context of growth (McKane et al., 2007; Martin et al., 2009) and may be closely related to the emergence of the phenomenon of life itself. The role of ion gradients and fluxes in the study of plant cell growth and elongation in plants, especially calcium and protons in the case of pollen tubes, has attracted a great deal of attention (Holdaway-Clarke and Hepler, 2003, for review). Calcium is one of most abundant elements in the earth's crust and plays a wide variety of roles within cells (Martins et al., 2013) – calcium ion concentrations can lead to localised or even plant-wide oscillations that can regulate downstream events. Cells maintain a steady state $Ca^{2+}$ concentration about hundred times greater than its concentration in the extracellular space. Such a large difference in concentration allows for rapid oscillations through the opening of calcium channels in the separating membranes. For our purposes, it is important to note that "gating" permits the maintenance of large differences in concentration with the exterior of the cell, as well as in the orchestrated release of calcium during signalling events. $H^+$-ATPases generate a voltage of about –150 mV across the plasma membranes of plants, which results in an electrochemical driving force of about 300 mV for cacium. A schematic diagram of the different mechanisms for the passive and active transport of $Ca^{2+}$ is presented in the review article by Martins et al. (2013, Fig. 2). The oscillations of calcium concentration open and close DMI1 channel, thus leading to a periodic $K^+$ ions efflux (ibid). On the other hand, Shabala et al. (1997) reported measurements of $Ca^{2+}$ and $H^+$ fluxes around the elongation region of corn roots at various pH levels. They noted that oscillations in ion fluxes were always detectable if the length of the observation was long enough. They also concluded that the different oscillatory components of ion fluxes around the roots of a plant appeared to be related to different ion transport systems. Investigations conducted by Zonia et al. (2002) indicated that the dynamics of the $Cl^-$ ion is an important ingredient in the network of events that regulate pollen tube homeostasis and growth. According to Zonia, oscillatory chloride efflux at the pollen tube apex plays a role in growth and cell volume regulation. By using an ion-specific vibrating probe the oscillatory growth of pollen tubes has been correlated with oscillatory influxes of the cations $Ca^{2+}$, $H^+$ and $K^+$.

By mentioning only these few reports from the abundant literature on ion transport being tightly coupled to growth, we have exemplified that oscillations (periodic motion) are ubiquitous in the elongation zones of growing plant organs. The latter statement is correct regardless of whether we are dealing with the multi-cell elongation zone in maize coleoptiles or roots, or the single-cell (subapical) elongating part of a pollen tube. Based on this observation, we propose a new experimental method, which allowed us to retrieve the basic phase relations and ion oscillation frequencies of an intact *Zea mays* L. plant in a single run of the experiment. The results that were obtained – ion frequencies, amplitudes (magnitudes) and phases – were compared with the available literature data for identification.

A key role for understanding the basic (both oscillatory and monotonic) growth mechanisms is attributed to ionic transport across membrane barrier. As was mentioned above, there are countless articles on passive and active transport in plants and the influx and efflux of substances, such as sucrose or hexose or different anions and cations, have been described in detail. Many trials have been undertaken to systematise this knowledge not only on the level of plant kingdom or species, but also for organs, tissues, cells or organelles. One of the most famous is the "Overview of the various transport processes on the plasma membrane and tonoplast of the plant cells", which is presented in Fig. 6.14 in Taiz and Zeiger (2006) although even here there is no information about the time sequence of the events (processes) in the cell cycle. However, it is highly credible that the time sequence may be fundamental for finding the solution to many discussions among biologists about the most basic mechanisms, such as the role of turgor pressure in the building of the primary wall. In this context, our proposal of a new universal and relatively inexpensive method, which is able to detect and discriminate the frequency landscape of a growing plant, seems to be a milestone in growth physiology. It is clear that the presented method, although not completely autonomous, is drastically limited by the available literature on ionic transport in plants, and the fact that most research is conducted on model plants.

This work presents the very subtle electrical measurements of growing maize coleoptiles that were obtained using an external (non-invasive) EMF measuring system. The data were compared with other oscillatory electrical and ionic data from other systems and were subjected to frequency spectrum analysis. The results are essentially correlative and provide information about the timing of underlying processes.

**Materials and methods**

*Investigation of $Ca^{2+}$, $H^+$, $K^+$ and $Cl^-$ fluxes in single maize coleoptile in situ*

Maize cells were used as the experimental system with which to study ion transport processes and establish the characteristic timing in the elongating cell complex an individual higher plant using solenoid-based circuit. Measurements were carried out on multi-cell organs (three-day-old coleoptiles) of maize (*Zea mays* L.) *in situ*. Seeds of maize were grown on a Hoagland's medium (Hoagland and Arnon, 1950) and cultivated in darkness at 27 ºC. Then they were placed (*in vivo* studies) in artificial pond water (APW) in a dark chamber (Supplementary Information Figure S1) at a fairly constant temperature. The temperature was measured before and after the experiment (24 – 25 ºC in which the optimum growth of maize takes place). The pre-incubation time was 30 min., while the duration of the experiment was from one to twenty hours (a four-hour experiment, which was chosen as the most efficient one after many trials, is presented here). The thermally isolated measuring chamber was shielded from the electromagnetic fields of environment. It was covered with Al thin foil to create a Faraday cage. Both the chamber and the 6 ½ Digit Precision Multimeter were grounded. The initial length of the seedling and the length increment was read off. In order to optimise the measurement conditions, we performed 39 experiments of different durations. In almost all of the experiments the coil was placed about 3 – 5 mm below the tip at the beginning of the experiment (Figure S2). The fragment of maize coleoptiles that was investigated elongated most intensely and was free from cell divisions. Ice was added into an additional container (with an immersed NiCr-NiAl thermocouple) just before the measurement, so that it wouldn't melt in the course of experiment (Figure S3). The measurement was usually done after several minutes of the primary stabilisation of the whole system. The growing seedlings remained untouched during this non-invasive experiment and were able to continue to grow and develop leaves after the experiment (Figure S2).

Both, the co-moving (with the advancing tip) and the fixed reference system (located at the shank) were used. In the case of co-moving frame, the increasing distance of the coil form the tip allowed unique continuous measurements to be performed at the three apical and subapical zones (tip, subapex – intermediate zone, shank) of the same growing sample. In our method, we also omitted the technically demanding problem related to the difficulty of impaling the apex in order to perform the measurements, while still maintaining elongation.

*DFT analysis*

The (negative feedback) RL circuit was formed tightly in the form of a solenoid on the growing coleoptile (Figure S2). The Cu wires in the form of a shielded twisted pair were used to connect to the rest of the system. The net voltage (thermocouple 'constant' voltage V plus Lenz-rule – induced $V_{EMF}$ stating that the induced EMF and induced current were in such a direction as to oppose the cause that produced them) was measured (Figures S3 – S4). The corresponding constitutive relation was used: $V_{EMF} = d\Phi/dt = d(LJ)/dt = LdJ/dt$. The inductance $L = 23.68$ µH was calculated from the following data: number of turns $n = 118$, length $0.75 \pm 0.01$ cm, radius $0.2 \pm 0.01$ cm and permeability of a vacuum $\mu = 4\pi \cdot 10^{-7}$ H/m. The coil Cu wire diameter equalled 0.12 mm, Cu connectors 0.16 mm; Cu plates that served as conductance electrons' reservoir (Matlak et al., 2001; Matlak and Pietruszka, 2004), with an average volume of $158.36 \pm 0.1$ mm$^3$. The absorption spectrum measurement was possible not only because of the long duration of the experiment (3600 s up to 72000 s) with dense (1 s) sampling, but presumably it was also amplified due to the type of "collective excitation" of many (~ $0.5 \cdot 10^4$, Berg et al., 2009) cells during one cycle. Note, that the synchronicity of different coupled oscillators may, by preventing destructive interference, increase the robustness of the signal (Kroeger and Geitmann, 2013). By assuming typical values for $dJ \sim 1$ pA and $dt \sim 1$ ms, we got $V_{EMF} \sim 10^{-13}$ V. However, by taking into account, by straightforward calculation, the number of cells in the investigated sample, and assuming about 100 channels per cell, we finally got the estimate $V_{EMF} = 10^{-6}$ V, which is within the reach of our measuring apparatus (µV). To be on the safe side, the measurements were performed in the ac and dc modes of the voltmeter and were recorded by a computer program (see Figure S5 for the time series). The 'void' measurements with an empty coil (with no coleoptile placed inside) were performed twice for reference purpose and the Fourier decomposition revealed no signal.

The absorption lines were then detected *via* the discrete Fourier transform (DFT), see Harris (1998) and Dieckmann (2011), and the read-off frequencies were associated and compared to the literature data (Tables S1 – S4, Figures S6 – S7). Recognition of these extremely small signals was possible because the duration of the experiment was long enough compared to the basic period of the investigated signal, and dense sampling. For example, a period $T$ of about 16.5 seconds (maximum of $Ca^{2+}$ influx) was encountered 873 times in a four-hour probe. If any regularly occurring event happened every 15 seconds within the measured time series, it would clearly be detected by the discrete Fourier transform to produce a spectral line that was connected with this event. The Fourier spectrum delivers information about the amplitudes, frequencies and phases of the basic cosines that span the function of which an interval that has a length of an integer number of periods is being sampled. Reconstruction of the original shape pulse function (time series) can be completed from a superposition of cosine functions with amplitudes and phases. The Fourier analysis technique is based on the fact that a signal can be decomposed into a sum of cosine or sine functions with periods that are a multiple of the total recording time. A certain periodicity in a recording is reflected by the large weight (Fourier coefficient) that multiplies the cosine or sine function with the respective periodicity in the Fourier decomposition (Fourier series). The value of these Fourier coefficients as a function of their period (or frequency) is called the Fourier transform. Since the Fourier transform is a complex quantity, the power of the Fourier series is usually calculated and is a real quantity (Zerzour et al., 2009). Here, we used the amplitudes, frequencies and phases as the basic ingredients of the spectral analysis (Dieckmann, 2011).

Note that the usual problem of the limited temporal resolution of data acquisition, which has hitherto prevented a more detailed analysis, was eliminated in our method by the dense sampling that we applied, by the use of an acquisition interval of 1 s, and the sufficiently long duration of the experiment (up to 72 000 measurements in a single run).

*System stabilization and cross-correlations*

System stabilisation, which was controlled by cross – correlations, is shown in Figure S8. The system that was investigated stabilised after the first hour of the measurement (Figures S8B and S8C). Closer examination revealed that the corresponding two time series produced, the auto-correlation triangle instead of the usual cross-correlation, which is presented in Figure S8D. This outcome delivers a convincing argument for the reproducibility of the results, thus supporting the proposed measurement method.

**Results**

*Frequencies and phases*

Fluxes of $H^+$, $K^+$, $Ca^{2+}$ and $Cl^-$ were measured at the elongating zone of growing maize coleoptile segment using negative feedback RL circuit (Figs S1 – S4). Discrete Fourier analysis of the time series that were obtained allowed the frequencies and phases of pulsatile ionic fluxes that were active during elongation growth, which are shown in Fig. S7, to be decoded. In order to categorise the type of ions that took part in the growth event, the data was deciphered Supplementary Information Tables S1 – S4 and compared with Fig. 3 in Holdaway-Clarke and Hepler (2003). The result that is based on this procedure is presented in Fig. 1.

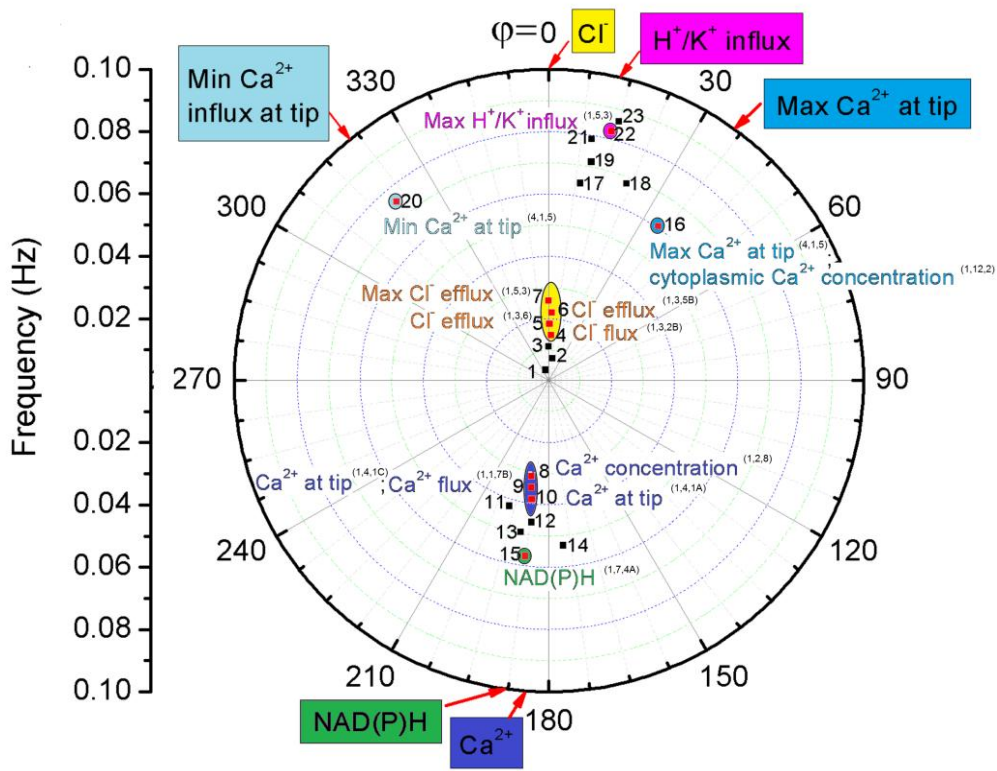

**Figure 1** Sequence of events in maize (*Zea mays* L.) growing tip as indicated in a "radar display" (polar coordinates) chart. The identified frequencies (square dots) – radial coordinate; phase offsets in degrees – angular coordinate (phase ϕ = 0 was assumed for $Cl^-$ efflux). Red squares – data identified in the literature; black squares – unidentified data, which may be associated with numerous transport processes such as those presented in Fig. 6.14 in Taiz and Zeiger (2006). Number coding by the upper indices: (SI table, reference, figure in the quoted article). Compare with Fig. 3 in Holdaway-Clarke and Hepler (2003).

**Amplitudes and spectrum**

DFT analysis also delivered the intensities of the processes (magnitudes) that describe the relative strength of the actual process and phase shifts. The amplitudes at given frequencies and the corresponding phases that were obtained *via* DFT analysis are presented in Fig. S7. The data displayed in Fig. S7 was assembled in a polar coordinates chart, Fig. 2. Some of this data was identified by a comparison with the literature data, which is collected in Tables S1 – S4. The resulting full spectrum of the oscillating ionic fluxes was created in Fig. 3 from the corresponding data points in Figs 1 – 2.

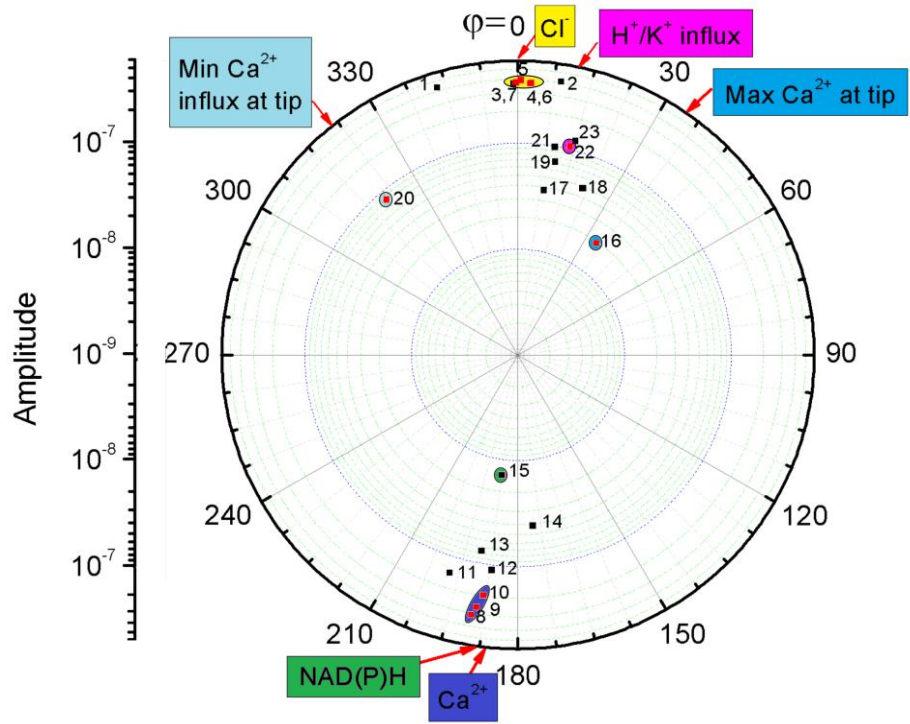

**Figure 2** Amplitudes (intensities) of maize (*Zea mays* L.) ionic fluxes in the growing tip presented as polar coordinates. Identified intensities (square dots) – radial coordinate; phase offsets in degrees – angular coordinate (phase φ = 0 is assumed for Cl⁻ efflux). Red squares – data identified in the literature; black squares – unidentified data, which, however, may be associated with the numerous transport processes that are presented in Fig. 6.14 in Taiz and Zeiger (2006). Number coding is the same as in Fig. 1.

Last but not least, the lowest lying frequencies in Fig. S9 are very close to the basic frequency $f = 0.066$ Hz that was obtained independently by theoretical and experimental methods for $Ca^{2+}$ oscillations for a lily in Fig. 5, as presented by Pietruszka and Haduch-Sendecka (2015). Moreover, this fundamental frequency is almost identical to the re-analysed data in Fig. S9 for lily pollen tubes and for that obtained for maize in this work, line $f$ = 0.064 Hz in Fig. S7. This may indicate that this lowest lying frequency for $Ca^{2+}$ is the same not only for these two species, but may even be common for other plants.

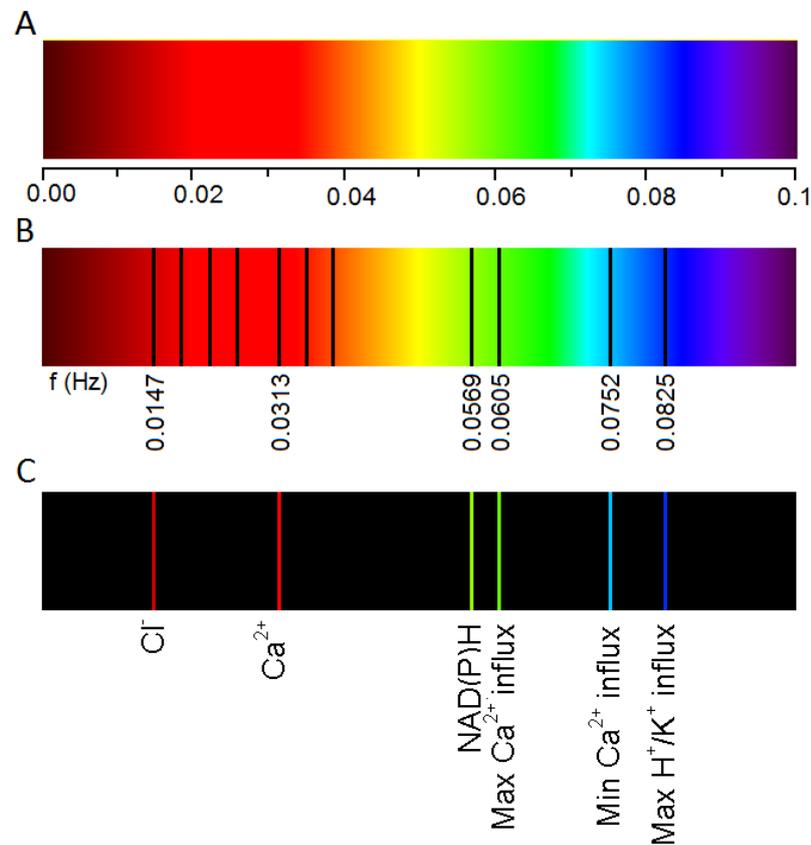

**Figure 3** Spectral landscape of a growing maize (*Zea mays* L.) coleoptile segment. **A.** Frequency scale in Hz. **B.** The absorption spectrum is divided into a number of spectral series (basic frequency values are indicated and the higher harmonics that were identified for $Cl^-$ and $Ca^{2+}$ are shown). Note the $Ca^{2+}$ base frequency that is slightly blue-shifted. **C.** Absorption spectrum of basic frequencies for the various ion fluxes that were obtained in our experiment and identified through a comparison with the literature data. Higher harmonics (recognised in the literature and collected in Tables S1 – S4) are shown in B: three lines above 0.0147 Hz for $Cl^-$ and two lines above 0.0313 Hz for $Ca^{2+}$.

**Conclusions**

Based on the proposed technique, a universal behaviour can be observed in the plant kingdom. Comparison with the literature data (Methods) revealed the existence of equal frequencies of the ionic channel fluxes either for growing single-cell pollen tubes or the multi-cell organs of higher plants. Moreover, phase shift analysis indicates that certain (ionic) processes occur with similar timings.

The new measurement method that is presented, though not entirely autonomous, was focused on finding a complete landscape (spectrum) of well-defined frequencies, intensities and phases of the oscillating ionic fluxes in plants. This turned out to be possible in a single run of a one-penny[*] experiment for an intact growing plant (*Zea mays* L.). Therefore, this method may be useful for the further analysis of the plant cell cycle, at least as a complementary technique to other well established methods (e.g. patch-clamp, vibration electrode) of the investigation of ionic currents in plant physiology, especially since we are well aware of how difficult it is to identify and characterise any single transport process. Further development of this experimental method may include simultaneous EMF and elongation (volumetric) growth measurements in order to interrelate the underlying microscopic processes with macroscopic (CCD camera) observations.

*) Copper bar India average price $9.0339/kg gives 0.1 cent/0.1 g solenoid (metalprices.com, 14.04.2015)

**Acknowledgements**

We thank E. Rówiński for technical assistance with Tektronix DMM 4040 6-1/2 Digit Precision Multimeter and P. Zajdel for the Python RS232/USB communication code (.py).

**Author contributions statement**

MP conceived the experimental method, MP conceived and wrote the paper, MP and AH-S constructed the experimental setup. AH-S and MP conducted the experiments, MP and AH-S analysed the results. Both authors reviewed the manuscript.

**Additional information**

The authors declare no competing financial interest.

**SI Tables and Figures Captions**

**Table S1** Ion influx and efflux events and corresponding time periods and frequencies in pollen tubes. *n* – denotes the number of data that were read off from the indicated source. *T* – stands for the oscillation period, while *f* denotes the corresponding frequency.

**Table S2** Ion influx and efflux events and corresponding time periods and frequencies in maize.

**Table S3** Data from Tables S1 and S2 that were used for analysis (*T* – period of oscillations).

**Table S4** Influx and efflux events and the corresponding phases and frequencies for a lily pollen tube.

**Figure S1** Experimental setup. Growing maize (*Zea mays* L.) coleoptile was placed in a thermally isolated (Styrofoam) Faraday cage (Al film) that was situated on a heavy marble plate in order to filter high-frequency ground oscillations.

**Figure S2** Maize (*Zea mays* L.) sample measurement *in situ.* During the experiment the solenoid was moved from the tip to a more basal position of a coleoptile. The Cu connectors (wires), which were connected to the experimental setup, are visible. **A.** The investigated sample that was prepared for the experiment – the solenoid is initially placed 5 mm below the tip of a three-day-old seedling. **B.** Photograph taken after a 24 hours measurement. The solenoid was moved to a position of 6.25 mm below the tip. The coleoptile elongation equalled 14.85 mm during the experiment.

**Figure S3** Schematic diagram of the measuring apparatus: Cu solenoid (inductance $L \sim 23$ μH), which was wound up tightly on the maize coleoptiles in the apical part, and was connected to the high precision digital multimeter (Tektronix DMM 4040 6 ½ Digit Precision Multimeter) *via* Cu square plates (Matlak and Pietruszka, 2004) and to the NiCr-NiAl thermocouple that was completely immersed in the ice container, which formed the parallel electric circuit. The voltage was measured and recorded through a RS232/USB interface and the data acquisition was completed on a computer using the Python code (.py).

**Figure S4** The solenoid that was used for measurements. The elongating zone of the coleoptile is shown in green. The figure is based on the Shipway and Shipway (2008) solenoid properties calculator.

**Figure S5** Electromotive force U (V) as a function of time (s). The original data points measured in the dc multimeter mode. The inset shows an oscillatory, step-wise signal structure.

**Figure S6** Logarithmic amplitude of a pulse spectrum. The measurement was performed below the coleoptile tip. A sample seed immersed in the APW. Experiment duration 3600 s; sampling 1s, number of solenoid turns n = 118. Compare with Dieckmann (2014).

**Figure S7** Logarithmic amplitude and phase angle of a pulse spectrum with the frequencies indicated.

**Figure S8** Logarithmic amplitude of a pulse spectrum and corresponding phases. Experiment stabilisation (steady state) is shown in the charts: DFT spectrum was calculated after (A) 1 hour (B) 2 hours (C) 3 hours. (D) Cross-correlations of B and C as a function of time lag (for DFT amplitudes – see main chart, and for the voltage time series – the inset). Cross-correlation is a measure of the similarity of two waveforms as a function of a time-lag that is applied to one of them. By definition, for continuous functions *f* and *g*, the cross-correlation is defined as**:**

$$(f * g)(\tau) \equiv \int_{-\infty}^{\infty} f^*(t) g(t+\tau) dt$$

where *f\** denotes the complex conjugate of *f* and $\tau$ is the time lag. In autocorrelation, which is the cross-correlation of a signal with itself, there will always be a peak at a lag of zero. Apparently, the cross-correlation of B and C seem to fulfil this definition (D), even though the experimental data originated from the subsequent (1 hour time delay) measurements. The latter result gives strong support to the experimental method that is introduced in this work.

**Figure S9** Linear interpolation $T_n = A + Bn$ of a basic period and higher harmonics for $Ca^{2+}$ signal for lily. Data based on Table S3 (1). $T_1 = 11.67 + 2.22 = 13.89$ s ($f_1 = 0.072$ Hz); $T_2 = 11.67 + 2*2.22 = 16.11$ s ($f_2 = 0.062$ Hz) – these lowest lying frequencies are close to the basic frequency $f = 0.066$ Hz that was obtained by Pietruszka and Haduch-Sendecka (2015) for the $Ca^{2+}$ oscillations that are shown in Figure 5.